\documentstyle[12pt,aasms4,psfig]{article}
\evensidemargin=\oddsidemargin

\def\gsim{\mathrel{\hbox{\rlap{\lower.55ex \hbox {$\sim$}}
                   \kern-.3em \raise.4ex \hbox{$>$}}}}

\def\hide#1{}

\begin{document}

\begin{center}{\bf 
Discovery of the first accretion-powered millisecond X-ray pulsar}

\vspace{1cm}

{Rudy Wijnands$^*$ \& Michiel van der Klis$^{*\dag}$}

\end{center}

\noindent $^*$ Astronomical Institute ``Anton Pannekoek'', University
of Amsterdam, \& Center for High Energy Astrophysics, Kruislaan 403,
1098 SJ Amsterdam, The Netherlands

\noindent $^\dag$ Department of Astronomy, University of California,
Berkeley, Berkeley, CA 94720
\vfill
\begin{center}Sent to Nature on April 20, 1998\end{center}
\vfill 
\newpage

\hide{max 180 woorden, bold part is nu 194 woorden, totaal text
(including bold part, but without all remarks and hides) is
1623. Criteria voor 2.5 pagina's in Natures style is 1500 worden, plus
4 figuren en/of tables.}

{\bf The precise origins of the millisecond radio pulsars, discovered
in the early 1980s$^1$, remain uncertain until this day. They
plausibly evolve from accreting low magnetic-field neutron stars in
X-ray binary systems$^{2,3}$. If so, these stars should spin at
millisecond rates.  In accordance with this idea, quasi-periodic
oscillations discovered in X-ray binaries around 50 Hz$^4$ and 1
kHz$^{5,6}$, and drifting oscillations at several 100 Hz in X-ray
bursts$^{6}$ have all been interpreted in terms of millisecond spins
of weakly magnetized neutron stars$^{7,8}$. However, in 15 years of
searching$^{9,10,11,12}$, the expected coherent millisecond signals
from X-ray binaries remained elusive.  In this Letter, we report the
discovery$^{13}$ of the first example of such a signal. Using the
Rossi X-ray Timing Explorer we find persistent 2.49 millisecond X-ray
pulsations in an X-ray binary, which we interpret to come from an
accretion-powered millisecond X-ray pulsar in the system. This is the
first known object of its kind. It is likely to switch on as a
millisecond radio pulsar when the accretion turns off completely. The
source is positionally coincident$^{14,13}$ with the known transient
X-ray burster SAX\,J1808.4--3658$^{15}$, which also makes this the
first X-ray pulsar which exhibits thermonuclear X-ray bursts.}

The transient X-ray burster SAX\,J1808.4--3658 was discovered in
September 1996 with the Wide Field Cameras onboard
BeppoSAX$^{15}$. Two thermonuclear X-ray bursts were observed, and the
source was classified as a low-mass X-ray binary (LMXB)
at a probable distance of 4 kpc$^{15}$. Recently, a transient source,
positionally coincident with SAX\,J1808.4--3658 to within a few
arcmin, was detected$^{14}$ with the Proportional Counter Array (PCA)
onboard the Rossi X-ray Timing Explorer (RXTE), showing the source was
again in outburst. The 2--10 keV flux on 1998 April 11 was
$1.5\times10^{-9}$ erg cm$^{-2}$ s$^{-1}$.

Data with a time resolution of 1/8192 s were taken with the PCA on
1998 April 11. The 2--60 keV count rate varied between 892 and 1007
counts/sec, including a $\sim$130 c/s background; no X-ray bursts were
observed. We computed a power spectrum of $\sim$3000 s of X-ray
data. The spectrum clearly shows the X-ray pulsations near a frequency
of 401 Hz (Fig. 1). Their amplitude was $\sim$4.1\% (rms) in the 2--60
keV range, with little dependence on photon energy.

We then divided the data into 128 second segments and determined the
pulsation frequency in each segment by an epoch folding technique. The
signal was present in each segment. During our initial 2-hr observing
span the pulsation frequency showed a smooth, approximately sinusoidal
$\sim$110-min variation between 400.986 and 401.022 Hz in the
satellite frame.  We corrected for this smooth variation and found we
could coherently fold the pulsations. Attributing the smooth frequency
variation to Doppler shifts, in part due to the satellite motion, we
concluded that the intrinsic coherence of the signal exceeded 10$^6$,
leaving no other interpretation than the presence of a millisecond
X-ray pulsar in the system$^{13}$. We confirmed the presence of the
signal in later observations taken between 1998 April 13--18.  The
Doppler shifts clearly show up in a dynamical power spectrum (Fig. 2).

\hide{Michiel, Finger et al. bepaalde de amplitude waarden (ze
detecteerde de hamonische) voor 1744 op 6.2+-0.6 \% en 1.4+-0.6 \% van
fundamental amplitude voor de 2de en 3de harmonische respectively.}

The folded pulse profile (Fig. 3) is near-sinusoidal, with amplitudes
of the harmonics at 2 and 3 times the 401 Hz frequency of $\sim$11 and
$\sim$4\% of the 401 Hz amplitude, respectively.  This is similar to
what is observed$^{16}$ in the 0.5 sec pulsar GRO\,J1744--28 and might
be related to both systems having a low inclination. It is possible
that the spin period is 4.98 ms with two near-identical pulses per
cycle. However, folding at this period indicates no significant
difference in shape between the odd and even pulses. The folded light
curves in the 2--5.0, 5.0--8.6 and 8.6--60 keV photon energy bands are
not significantly different.

Below 100 Hz the power spectrum somewhat resembles that of other
accretion-powered X-ray pulsars as well as of low-luminosity
LMXBs$^{17}$. There is a strong ($\sim$20\% rms, 2--60 keV) broad
noise component below 100 Hz, with a bump near 12 Hz that, when
interpreted in terms of the magnetospheric beat-frequency model$^7$
would suggest an inner disk radius of 30 km (just inside the
corotation radius). An additional $\sim$17\% (rms) noise component
extends out to roughly the 0.4 kHz pulse frequency. This is {\it not}
commonly seen in LMXBs, but is, at proportionally lower cutoff
frequencies, in other accreting pulsars$^{17}$.  No kilohertz
quasi-periodic oscillations$^{18}$ are detected. However, upper limits
(3--6\% rms), when compared to observed amplitudes in other LMXBs, are
as yet insufficient to exclude their presence.

The detection of the pulsations from an X-ray transient that,
moreover, likely showed thermonuclear X-ray bursts during an earlier
outburst, identify this X-ray pulsar as one that is powered by
accretion from a binary companion, and pulses because magnetically
channeled accretion produces hot spots on the neutron star surface
which spin around with the star's spin.  In a subsequent analysis
Chakrabarty and Morgan$^{19,20}$ have found that, correcting for the
satellite orbit, the pulsar's Doppler shifts indicate it is in a 2-hr
orbit around a low-mass star, with an intrinsic pulse frequency of
400.9753(1) Hz$^{19}$.

For accretion to proceed uninhibited by the centrifugal force on the
accreting matter corotating in the magnetosphere, the magnetic field
strength $B$ of this pulsar must be much lower than the
$\sim$10$^{12}$ Gauss typical for slower X-ray pulsars. Centrifugal
inhibition occurs when the magnetospheric radius $r_M = {\rm 18 km}\,
\xi \mu_{26}^{4/7} m^{1/7} R_{6}^{-2/7} L_{37}^{-2/7}$ is larger than
the corotation radius $r_{co} = {\rm 15 km}\, m^{1/3} P_{\rm 1
ms}^{2/3}$, where $\mu_{26}$ is the magnetic moment in 10$^{26}$ Gauss
cm$^3$, $L_{37}$ the bolometric luminosity in 10$^{37}$ erg/s, $m$ the
neutron star mass in solar masses, $R_6$ its radius in units of $10^6$
cm, $\xi$ a dimensionless factor that is probably between 0.5 and
1$^{21}$, and $P_{\rm 1 ms}$ the spin period in ms. On 1998 April 11
the 2--10 keV luminosity, assuming a distance$^{15}$ of 4 kpc, was
$\sim3\times10^{36}$ erg s$^{-1}$. The spectrum is quite hard$^{22}$,
and the bolometric luminosity may have been closer to $6\times10^{36}$
erg s$^{-1}$. With the above expressions this leads to an upper limit
on $B$ of $\sim$2--6$\times 10^8$ Gauss (4--14$\times 10^8$ Gauss if the
pulse period is 4.98 ms). As the X-ray flux decreases in the transient
decay this upper limit will become less. When centrifugal inhibition
sets in we might see a rapid drop in X-ray flux$^{23,24}$; this would
allow to estimate the value of $B$.

Such a low magnetic field strength is in the range of values inferred
for millisecond radio pulsar$^{25}$. The object appears to be located
above the pulsar ``death line''$^{25}$, so when accretion finally
turns off, this source would likely switch on as a radio pulsar. This
might even happen {\it between} X-ray outbursts, if the density in the
pulsar environment, at least out to the light cylinder at 120 km,
drops sufficiently. So, this low-mass X-ray binary could indeed be one
of the progenitors of the millisecond radio pulsars, and it might be
an intermittent radio pulsar now.

That this pulsar likely also bursts is in accordance with theoretical
expectations. Normally, X-ray bursts do not occur in accreting
pulsars, as due to the magnetic field either the nuclear burning takes
place steadily$^{26}$, or nuclear burning fronts can not propagate
rapidly$^{27}$. However, if the field is weak, these mechanisms are
ineffective and bursts could occur$^{21}$. This seems to be verified
by SAX\,J1808.4--3658. This is the first time pulsations and
thermonuclear burst are seen in one source. The bursts in the 0.5 s
pulsar GRO\,J1744--28$^{16,28}$ are not thermonuclear but due to
accretion instabilities$^{28,29}$.

As the likely $B$ field of SAX\,J1808.4--3658 is not much different
from that inferred for other LMXBs from models for the rapid aperiodic
variability$^{7}$ and the X-ray spectra$^{30}$, and its spin rate is
similar to that inferred in other LMXBs from models for burst
oscillations$^6$ and kHz QPOs$^{8}$ the question arises: Why is
SAX\,J1808.4--3658 the only known LMXB with a millisecond pulsar?
While it can not be excluded that at higher luminosity
SAX\,J1808.4--3658, like the other LMXBs, would not pulse, for LMXBs
of similar luminosity the simplest explanation would seem to be that
their $B$ fields are considerably weaker than that of this pulsar.  It
would be of great interest to detect in SAX\,J1808.4--3658 the
drifting oscillations during X-ray bursts interpreted in other LMXBs
as due to the neutron star spin$^6$, or the twin kHz QPO peaks whose
frequency difference has also been interpreted as the neutron star
spin frequency$^{5,6,8}$, as the fact that the spin frequency of
SAX\,J1808.4--3658 is known will provide a direct test of these
interpretations.

We finally note that the high-precision timing measurements possible
due to the high frequency of these pulsations will allow to study the
torques of the accreting matter on the neutron star and the star's
response to these torques with unprecedented accuracy. Moreover, for
the first time this will be possible in an accretion flow that extends
freely down to only a few tens of kilometers from the neutron star
surface. It is likely that this will provide new insights into the
nature of the inner accretion flows around weakly magnetized compact
objects. It may also provide new information about the internal
constitution of neutron stars.

\references

1. Backer, D.C., Kulkarni, S.R., Heiles, C., Davis, M.M., \& Goss,
W.M. A millisecond pulsar. {\it Nature} {\bf 300}, 615 (1982)\\
2. Srinivasan, G. \& van den Heuvel, E.P.J. Some constraints on the
evolutionary history of the binary pulsar PSR 1913+16. {\it Astr. Astrophys.}
{\bf 108} 143 (1982)\\
3. Alpar, M.A.,  Cheng, A.F., Ruderman, M.A., \& Shaham, J. A new class
of radio pulsars. {\it Nature} {\bf 300}, 728 (1982)\\
4. van der Klis, M., {\it et al.} Intensity-dependent quasi-periodic 
oscillations inthe X-ray flux of GX 5$-$1. {\it Nature} {\bf 316}, 225 (1985)\\
5. van der Klis, M. {\it et al.} Discovery of submillisecond
quasi-periodic oscillations in the X-ray flux of Scorpius X-1. {\it
Astrophys. J. Lett.} {\bf 469}, L1 (1996)\\
6. Strohmayer, T. E. {\it et al.} Millisecond X-ray variability from
an accreting neutron star system. {\it Astrophys. J. Lett.} {\bf 469},
L9 (1996)\\
7. Alpar, M.A. \& Shaham, J. Is GX 5--1 a millisecond pulsar? {\it
Nature} {\bf 316}, 239 (1985)\\
8. Miller, M.C., Lamb, F.K., \& Psaltis, D. Sonic-point model of
kilohertz QPOs in LMXBs. {\it Astrophys. J.} Submitted (1998)\\
9. Leahy, D.A. {\it et al.} On searches for pulsed emission with application
to four globular cluster X-ray sources: NGC 1851, 6441, 6624, and
6712. {\it Astrophys. J.} {\bf 266}, 160 (1983)\\
10. Mereghetti, S. \& Grindlay, J.E. A search for millisecond periodic
and quasi-periodic pulsations in low-mass X-ray binaries. {\it
Astrophys. J.} {\bf 312}, 727 (1987)\\
11. Wood, K.S. {\it et al.} Searches for millisecond pulsations in
low-mass X-ray binaries. {\it Astrophys. J.} {\bf 379}, 295 (1991)\\
12. Vaughan, B.A. {\it et al.} Searches for millisecond pulsations in
low-mass X-ray binaries, II. {\it Astrophys. J.} {\bf 435}, 362
(1994) \\
13. Wijnands, R. \& van der Klis, M. SAX\,J1808.4--3658 =
XTE\,J1808-369. {\it IAU Circ.} No. 6876 (1998)\\
14. Marshall, F.E. SAX\,J1808.4--3658 = XTE\,J1808--369. {\bf IAU Circ.} 
No. 6876 (1998)\\
15. in 't Zand, J.J.M. {\it et al.} Discovery of the X-ray transient
SAX\,J1808.4--3658, a likely low-mass X-ray binary. {\it
Astr. Astrophys. Lett.} 331, L25 (1998)\\
16. Finger, M.H. {\it et al.} Discovery of hard X-ray pulsations from
the transient source GRO\,J1744--28. {\it Nature} {\bf 381}, 291
(1996) \\
17. van der Klis, M. Rapid aperiodic variability in X-ray binaries. In
{\it X-ray binaries} (eds. Lewin, W.H.G., van Paradijs, J., \& van den
Heuvel, E.P.J.) 252 (University Press, Cambridge) (1995)\\
18. van der Klis, M. Kilohertz quasi-periodic oscillations in low-mass
X-ray binaries - a review. In {\it The many faces of neutron stars}
(eds. Buccheri, R., van Paradijs, J., \& Alpar, M.A.) (Kluwer Academic
Publisher) in press (1998)\\
19. Chakrabarty, D. \& Morgan, E.H. SAX\,J1808.4--3658 = XTE\,J1808--369.
{\it IAU Circ.} No. 6877 (1998)\\
20. Chakrabarty, D. \& Morgan, E.H. The 2 hour orbit of the bursting
millisecond X-ray pulsar SAX\,J1808.4--3658. {\it Nature} submitted (1998)\\
21. Bildsten, L. \& Brown, E. F. Thermonuclear burning on the
accreting X-ray pulsar GRO\,J1744--28. {\it Astrophys. J.} {\bf 477},
897 (1997) \\
22. Heindl, W., Marsden, D. \& Blanco, P. SAX\,J1808.4--3658 =
XTE\,J1808--369. {\it IAU Circ.} No. 6878 (1998) \\
23. Zhang, S.N., Yu, W., \& Zhang, W. Spectral state transition in
Aquila X-1: evidence for ``propeller'' effects. {\it
Astrophys. J. Lett.} {\bf 494}, L71 (1998) \\
24. Campana, S. {\it et al.} Aquila X-1 from outburst to quiescence:
the onset of the propeller effect and signs of a turned-on
rotation-powered pulsar. {\it Astrophys. J. Lett.} in press (1998)\\
25. Bhattacharya, D. \& van den Heuvel, E.P.J. Formation and evolution
of binary and millisecond radio pulsars. {\it Physics Reports}
{\bf 203}, 1 (1991)\\
26. Joss, P.C. \& Li, F.K. Helium-burning flashes on accreting neutron
stars: Effects of stellar mass radius, and magnetic field. {\it
Astrophys. J.} {\bf 238}, 287 (1980)\\
27. Bildsten, L. Propagation of nuclear burning fronts on accreting
neutron stars: X-ray bursts and sub-hertz noise. {\it Astrophys. J.} 
{\bf 438}, 852 (1995)\\
28. Kouveliotou, C. {\it et al.} A new type of transient high-energy
source in the direction of the Galactic Centre. {\it Nature} {\bf
379}, 799 (1996) \\
29.  Lewin, W.H.G., Rutledge, R.E., Kommers, J.M., van Paradijs, J.,
\& Kouveliotou, C. A comparison between the Rapid Burster and
GRO\,J1744--28. {\it Astrophys. J. Lett.} {\bf 462}, L39 (1996) \\ 
30. Psaltis, D., Lamb, F.K., \& Miller, G.S. X-ray spectra of Z sources. ({\it
Astrophys. J. Lett.} {\bf 454}, L137 (1995)\\



{\bf Acknowledgements.} We thank Lars Bildsten and Jan van Paradijs
for carefully reading the manuscript, them and Jon Arons for
stimulating discussions, and the RXTE team for promptly making the
data available online. MK gratefully acknowledges the Visiting Miller
Professor Program of the Miller Institute for Basic Research in
Science (UCB). We thank the Netherlands Organization for Research in
Astronomy ASTRON for financial support.

\newpage

\begin{figure}
$$\psfig{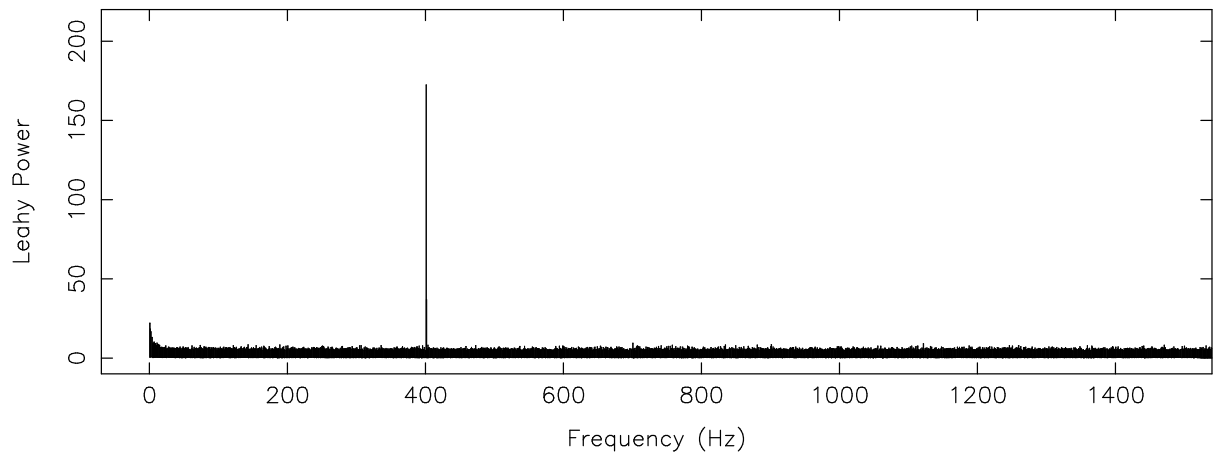}$$
\caption[]{Leahy$^9$ normalized power spectrum of the 1998 April 11
20:38--21:21 UT persistent emission of SAX\,J1808.4--3658. The
periodicity near 401 Hz is obvious. No harmonics or subharmonics are
seen.
\label{fig1}}
\end{figure}  

\begin{figure}
$$\psfig{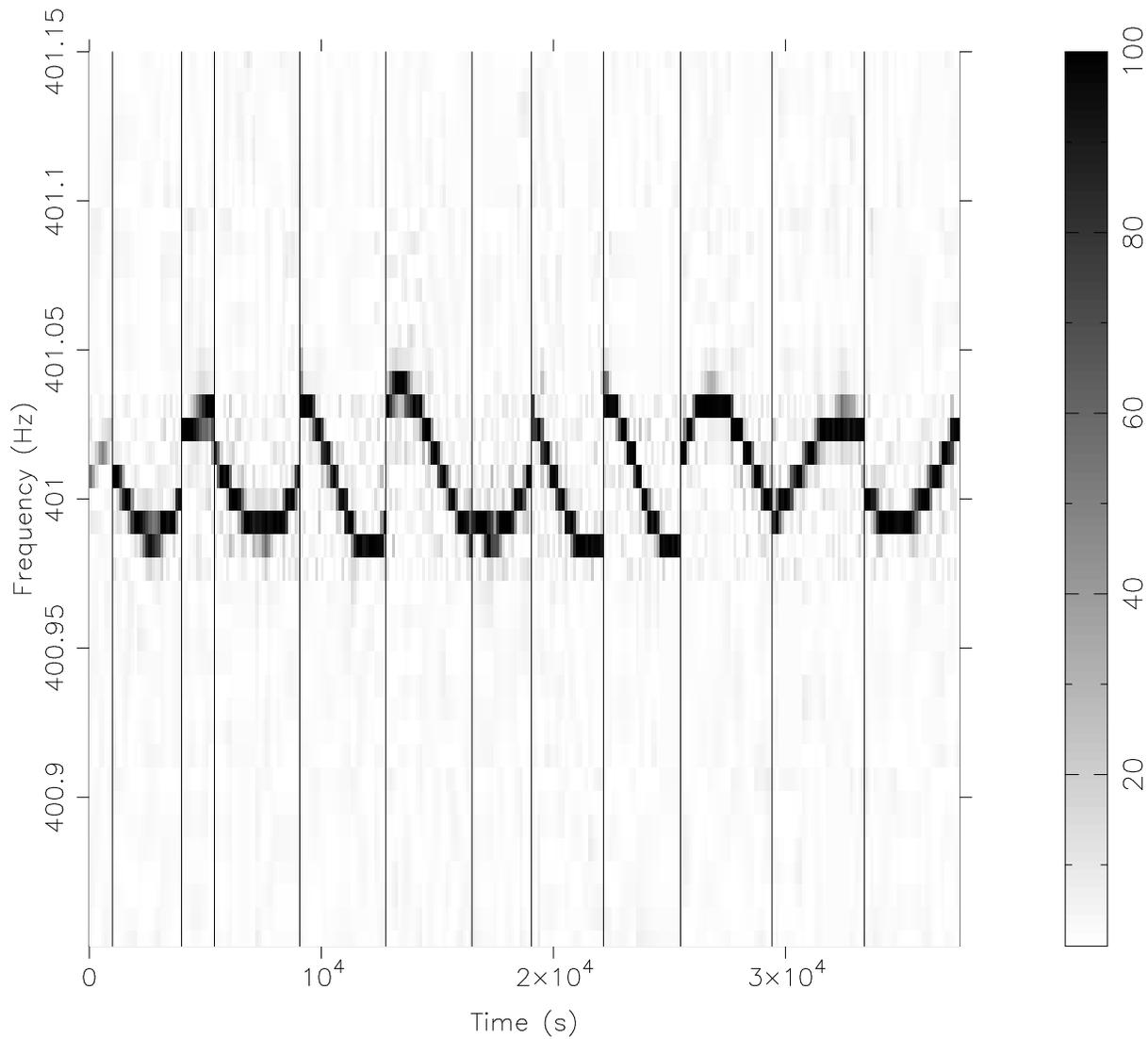}$$
\caption[]{Dynamical power spectrum of the 1998 April 11--18
data. The Doppler shifts in pulsation frequency due to the satellite
and pulsar orbital motion are evident. Time axis is not contiguous;
vertical lines indicate data gaps.}
\end{figure}

\begin{figure}
$$\psfig{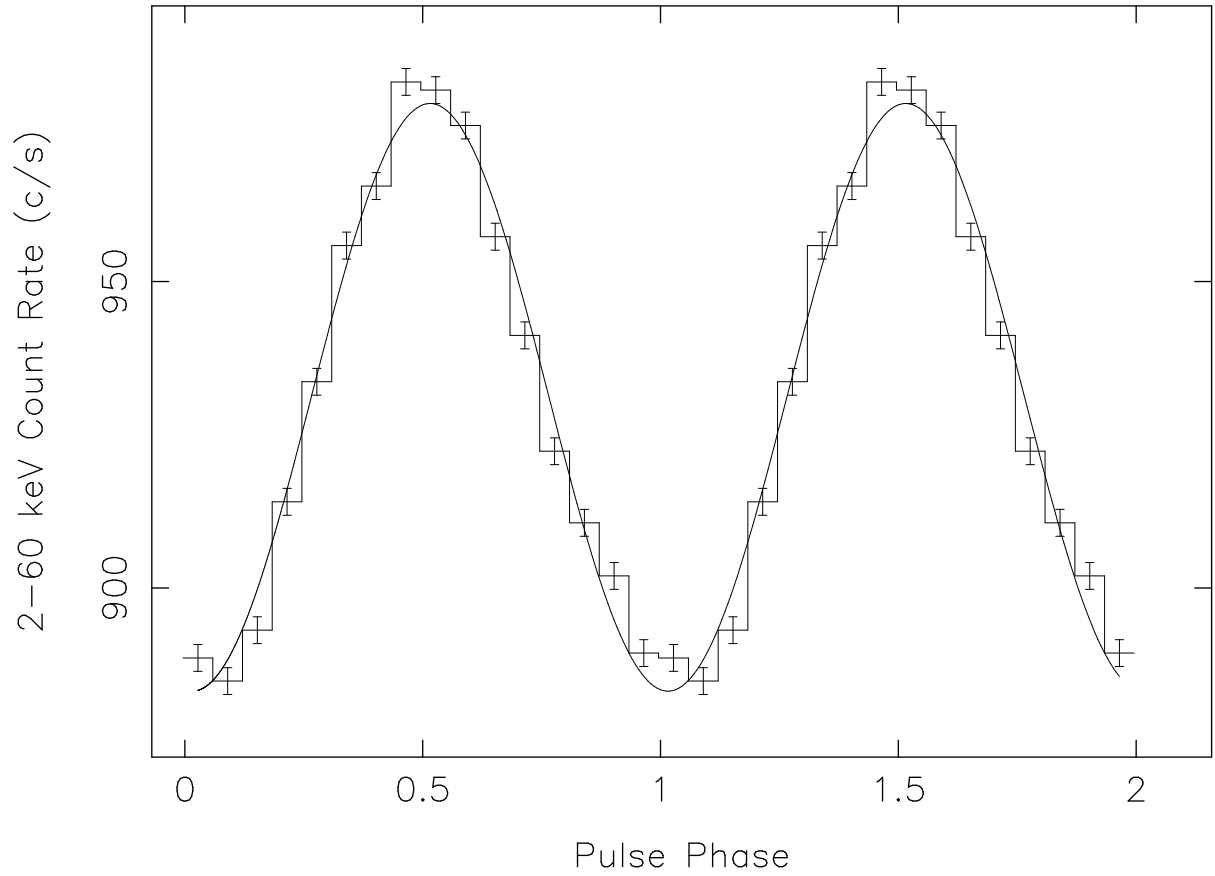}$$
\caption[]{The 2--60 keV light curve folded at the 2.49 ms period. Two
cycles are plotted for clarity. The solid line is the best fit
sinusoid.}
\end{figure}

\end{document}